\documentclass[conference]{IEEEtran}

\usepackage{cite}
\usepackage{graphicx}
\usepackage[tight,footnotesize]{subfigure}

\begin{document}
%
\title{Understanding periodicity and regularity of nodal encounters in mobile networks: A spectral analysis}

\author{\IEEEauthorblockN{Sungwook Moon, Ahmed Helmy}
\IEEEauthorblockA{Department of Computer and Information Science and Engineering\\ University of Florida\\
\{sungwook, helmy\}@ufl.edu} }

\maketitle

\begin{abstract}
Study on human mobility is gaining increasing attention from the research community with its multiple applications to use in mobile networks, particularly for the purpose of message delivery in the Delay Tolerant Networks. To better understand the potential of mobile nodes as message relays, our study investigates the encounter pattern of mobile devices. Specifically, we examine the extensive network traces that reflect mobility of communication devices. We analyze the periodicity in encounter pattern by using power spectral analysis. Strong periodicity was observed among rarely encountering mobile nodes while the periodicity was weaker among frequently encountering nodes. Further, we present a method to search regularly encountering pairs and discuss the findings. To our knowledge, we are the first to analyze the periodicity of encounter pattern with large network traces, which is a critical basis for designing an efficient delivery scheme using mobile nodes.
  
\end{abstract}

\IEEEpeerreviewmaketitle

\section{Introduction}
   Mobility and nodal encounters are utilized to deliver messages in intermittently connected delay tolerant networks (DTNs) \cite{dtn}. Much of DTN research so far has been devoted to the study of message delivery protocols and the design of mobility models. While these studies are essential for eventual implementation of the mobile adhoc networks using DTN concept, understanding of nodal encounter pattern is a critical basis for the success of protocol deployment as delivery mechanism depends on nodal encounter. It is worth to note the meaning of encounter in the networks. As our focus is on mobile networks, the term encounter in this work indicates the event that two or more users present within the wireless communication range. The terms, encounter and contact, are used interchangeably in literatures \cite{mining} \cite{powerlaw} and we use the term encounter throughout the paper for consistency. In this presentation, we explore the periodicity presences in encounter patterns and analyze them. By the spectral analysis of encounter pattern, we find the periodic patterns that are repetitive and discuss their applications later. To achieve this goal of study, we analyzed various types of the Wireless LAN (WLAN) and Bluetooth encounter traces. First, we generate the encounter traces with a reasonable assumption from WLAN traces. Bluetooth traces are naturally encounter traces without the need of any transformation as they log the identification of the Bluetooth device that the subject Bluetooth devices have discovered. However, the scale of Bluetooth traces are limited to the number of subjects carrying the devcies with the discover program on. Hence, we use the WLAN traces for scalable analysis of encounter pattern. In order to use the WLAN trace as encounter trace, we use common assumption that had been used by other publications \cite{mining} \cite{chaintreau}, which defines the encounter occurrence in the WLAN environment as the nodes that are associated with the same access points (APs) in the same period of time. After transformation to the encounter trace, the next step is generating a time series data for the number of metrics, namely, daily/hourly encounter, encounter frequency and encounter duration. We apply the Auto Correlation Function (ACF) to identify the repetitive patterns and perform power spectral analysis to find the distinct periodicities in encounter patterns for each metric. Fast Fourier Transform (FFT) was performed in conversion to frequency domain for computation efficiency and analyze the frequency magnitude in the spectrum. We highlight the important periodicities by differenct groups and discuss the utilization of the result in the mobile networks. After analyzing the periodicity, we show some of appraoches to extract the periodically encountering node pairs and conclude with the summary and applications.

   In the following section, we introduce the methodology to analyze the periodicity along with the encounter traces in section II. Analysis of periodicity for the encountered pairs and individual encounter patterns in WLAN and bluetooth traces follow in section III. Then, section IV describes the approaches to extract regularly encountering node pairs and discuss the results. We explain about related works in section V, and wrap up with conclusions and summary in section VI.


\section{Methodology}
   For spectral analysis of the encounter traces, multiple steps are required. In our work, raw network traces are processed to the encounter traces in the form of time series data. We apply the autocorrelation function (ACF), then transform them to the frequency domain by performing the discrete-time Fourier transform (DFT).  

\subsection{Encounter traces in the mobile networks}

   We use the two types of data sets for nodal encounter: Bluetooth traces and WLAN traces. Bluetooth traces reflect the encounters of users carrying mobile devices with Bluetooth communication capabilities. The limitation of Bluetooth traces is the scalability, as their data is collected by the participants willing to run the Bluetooth devices discovery program. Whereas, WLAN traces can be very large as the centralized system can continuously collect the data via access points belonging to the particular organization (e.g. college campus). As discussed earlier, we use an assumption that the users who accessed to the same access points (APs) have encounter events. This assumption may not reflect the exact encounter; however, it is close to real encounter considering the users accessed the same APs were at the close proximity of each other and could have communicated each other through the AP.

\begin{table}
    \begin{tabular}{ | p{1.2cm} | p{2.2cm} | p{1.2cm} | p{1cm} | p{1.2cm} |}
    \hline
    Trace source & Trace duration & Analyzed duration & Unique users & Encounter pairs \\ \hline \hline
    USC & 2006 Jan-May & 128 days & 28173 & 25359454 \\
	& 2007 Jan-May & & 35274 & 19057089 \\
	& 2008 Jan-May & & 42587 & 31289100 \\ \hline
    UF & 2007 Aug-Dec & 128 days & 46115 & 12493403 \\
       & 2008 Jan-May & & 50549 & 16807427 \\ \hline
    Monteral & 2004 Aug-Dec & 128 days & 455 & 2512 \\ \hline
    Bluetooth & 2/25 - 3/7/2008 & 256 hours & 10 & 1277 \\
	& 11/17 - 27/2008 & & 27 & 1655 \\  \hline
   \end{tabular}
	\caption{Statistics of encounter traces}
	\label{trace_table}
\end{table}

\subsubsection{Bluetooth encounter}
   Scale of Bluetooth encounter data is considerably small, compared to WLAN traces due to the difficulty of finding subjects to participate. Some of the available Bluetooth encounter data include the conference encounter \cite{conf} and bus encounter \cite{bus} \cite{haggle}. While these data sets may be useful for particular scenarios, we conducted our own experiment to observe general Bluetooth encounter, which matches to the WLAN trace we also collect. Each of graduate student taking the Computer Networking course in 2008 was assigned a PDA (HP iPAQ or Nokia N800/810) and was strongly encouraged to carry the mobile device as often as possible with the Bluetooth encounter collection program running. This program broadcasts the beacon signal every 60 seconds and logs the Bluetooth device information that acknowledges the beacon signal, including the timestamp. This experiment was performed for two semesters (2008 spring and fall) \cite{nile}, each with different groups of students. Due to the short length of experiment, we observed hourly encounter instead of daily encounter. As Table \ref{trace_table} shows, there are 10 and 27 subjects in spring semester and fall semester respectively. These collected Bluetooth traces contain the information of the encountered nodes, namely their MAC addresses and timestamps for acknowledgements. 

\subsubsection{WLAN traces}
   There are many forms of network traces available in public, which can be obtained from \cite{crawdad}, including the city of Montreal trace \cite{crawdad} that we use in this paper. To obtain large scale network traces that cover the entire campus over a length of more than one academic semester, we collected campus-wide WLAN traces at the University of Florida (UF, 2007 fall - 2008 spring semester) \cite{nile}. We also used the WLAN traces of the University of Southern California (USC, 2006-2009 spring semesters) \cite{nile} as in Fig. \ref{trace_table}. These WLAN traces have the following information - MAC address, associated AP and timestamp for start and end time of association. Based on the assumption described earlier, the WLAN traces are processed to encounter traces that log the MAC addresses of the encountered WLAN devices, timestamps, durations and locations of encounter. Conversion to the encounter trace is computation and storage consuming process. Given $m$ inputs in average for $n$ number of nodes, the computation needed is $\frac{n(n-1)}{2}*m^2$ as it requires the comparison for each input between two nodes to determine the occurrence of encounter and its duration. Therefore, we obtain $O(n^2m^2)$ for overall compuation time in generating encounter trace and $O(n^2m)$ for the size of encounter trace data. If the orignal trace is sorted in time sequence, the computation reduces to $O(n^2m)$ because the comparison for the inputs of two nodes can be performed in sequence of proceeding time. For the size of very large data, which has at least over 28,000 nodes with the inputs ranging up to several megabytes for a monthly data, it is realistic to break down the entire trace by the certain periods for analysis purpose. We analyze the 128 days of data from each trace for the above reasons and consistency in comparison. Further, this specific time span roughly covers the entire semester for both campuses. 

\subsection{Spectral representation}
   To capture the multiple aspects of encounter behavior, we look into the following variables: encounter frequency $F_{d} (i,j)$, daily encounter $E_{d}(i,j)$, houly encounter $E_{h}(i,j)$ and duration of encounter $L_{d}(i,j)$, where given $n$ number of nodes, $(i,j)$ is the encountered node $i$ and $j ( 0 \leq i < n, 0 \leq j < n, i \neq j )$ and $d$ is a day $( 0\leq d < T )$, where $T$ is a total length of trace in days. Daily Encounter is a binary process. For each encounter pair of nodes $i$ and $j$ on day d, $E_{d} ( i,j )=1$ if at least one encounter event occurs in day $d$; otherwise, $E_{d}(i,j)=0$. Hourly encounter is the same process as the daily encounter except the time unit is an hour. Daily and hourly encounter are interval counting with an interval being a day and an hour respectively. We adapt this idea from timely-counting by Song et al. \cite{interval}. Encounter Frequency is number of encounters a day and denoted by $F_{d}(i,j)=\mu$, where $\mu$ is the total number of encounter events occurred in day $d$ and $0 \leq F_{d}(i,j)<24*60*60$. Encounter duration is another metric used and denoted by $L_{d}(i,j)=\xi$, where $\xi$ is the total duration of encounter event occurred in seconds in day $i$ and $0 \leq T_{d}(i,j)<24*60*60$. The result periodicity patterns for the encounter frequency and duration appear very similar to the patterns in daily encounter; thus, analysis and observances for these two metrics are identical.

   In both of the encounter traces, the timestamps log in seconds. To better observe the daily and hourly encounter characteristics, we process the trace data in days and in hours, respectively. As noted earlier, 128 days are the time spans for each of the trace in this analysis. This time spans are also beneficial to the use of Fast Fourier Transform (FFT) in frequency analysis because it requires the length of data to be the power of 2. Applying FFT for each semester data enables fast processing of massive encounter data and helps observing distinct characteristics in a finer granularity by preventing the seasonal effect (repeated behavior by each semester) from affecting the result.

   ACF is a measure of correlation between observations at different lags (distances) apart \cite{timebook}, thus providing insight into the stream of data. We use ACF to find the repeatitive periodical patterns from the processed time-domain representation of encounter traces. When lag $k=0$, it compares the data stream to itself, and autocorrelation is maximum, which results in Variance$\left( \delta \right)$. Given the mean $\lambda$ of a pair $(i,j)$, we calculate autocorrelation coefficients (autocoefficients) for the pair, $r_k(i,j)$ for each lag $k ( 1 \leq k < T)$, by computing the series of autocoefficients as in the following:
  
\begin{equation}
r_k (i,j) = \frac{\sum_{d=0}^{T-k}(E_d-\lambda)(E_{d+k}-\lambda)}{\sum_{d=0}^{T-1}(E_d-\lambda)^2}
\label{acf}
\end{equation}

\begin{figure}
\centering
\subfigure[rarely encountering pairs, $0.1 \leq D_{rate}<0.2$]{
\includegraphics[width=3in]{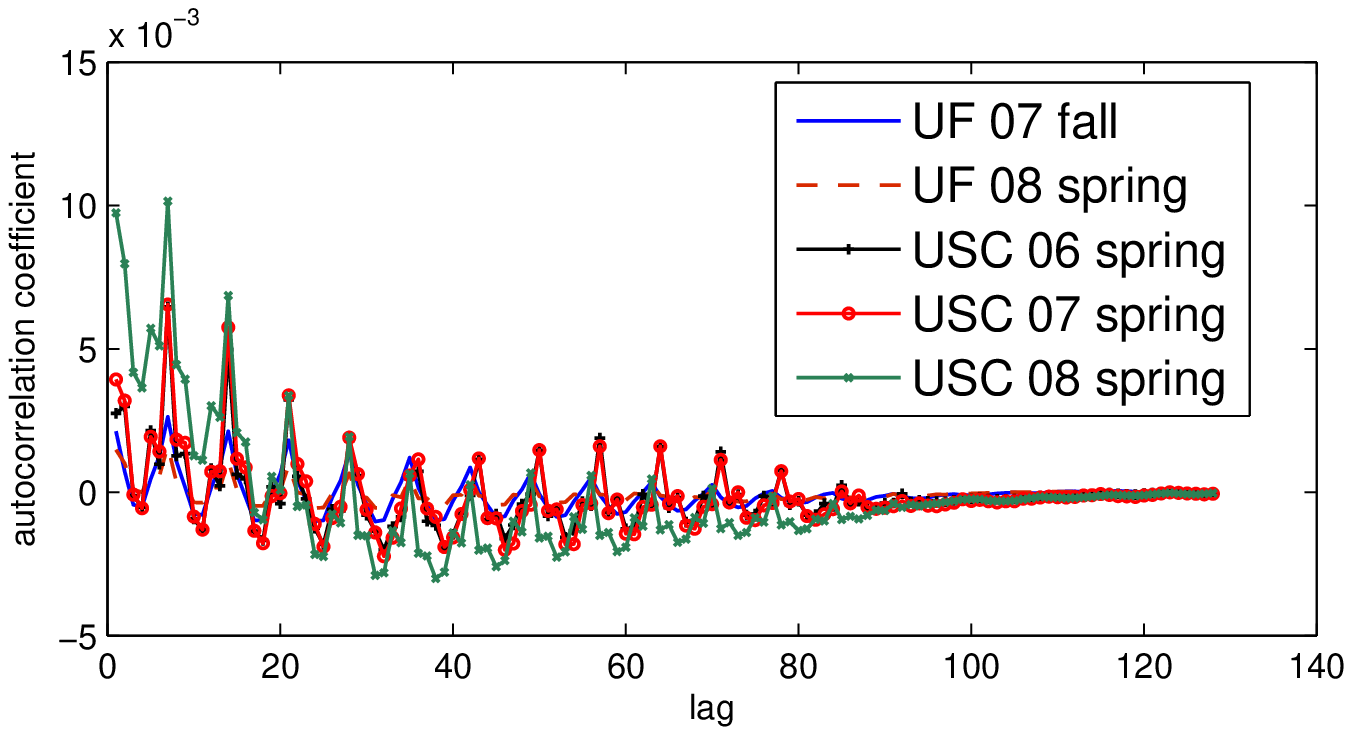} 
\label{fig_acf_rare}
}
\subfigure[frequently encountering pairs, $0.5 \leq D_{rate}<0.6$]{
\includegraphics[width=3in]{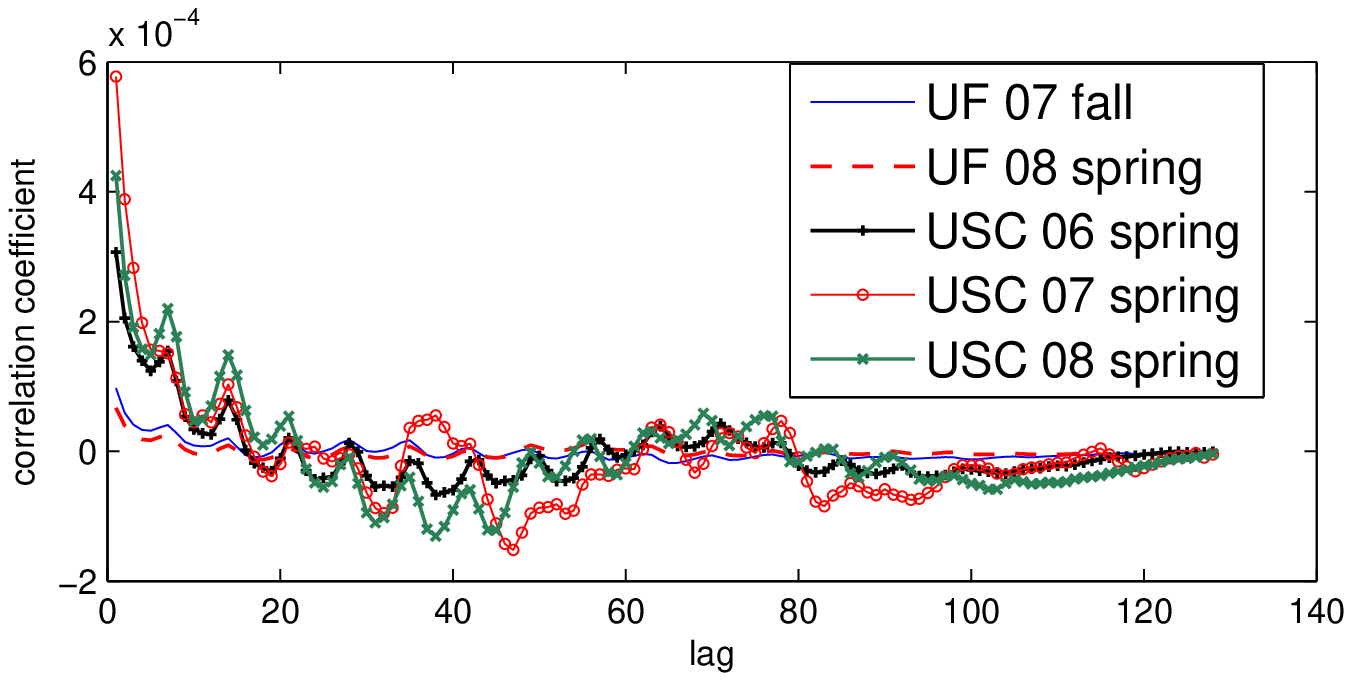}
\label{fig_acf_frq}
}
\caption{Average autocorrelation coefficient}
\label{fig_acf}
\end{figure}

   After producing the autocorrelation coefficients, we obtain a result showing periodic trends for certain lags as in Fig. \ref{fig_acf}. Yet, several distinct periodicities are hidden and require a further processing. Applying DFT is the necessary process in order to transform the autocoefficients of the time series data to the frequency domain. It produces the power spectrum $y_c$ of the pair $(i,j)$ for each frequency component $c (1 \leq c <T)$: 
\begin{equation}
y_c (i,j) = \sum_{k=1}^{T-1}r_k(i,j)e^{-\frac{2\pi i}{T}kc}
\label{fft}
\end{equation}
In resulted graphs, each bin in axis-X indicates the number of replicas over the observed period of time in frequency domain, while axis-Y is the frequency magnitude of corresponding frequency component for given trace length of $N$. In this representation, the sampling rate is $1/day$ for daily encounter and $1/hour$ for hourly encounter, naturally the Nyquist frequency becomes $0.5/day$ and $0.5/hour$ respectively.

\section{Periodicities in nodal encounters}

   To observe the periodicities of encountered pairs, we compare the average autocoefficient of each lag which are transformed to the frequency domain. 

\subsection{WLAN trace}
   Given the fact that the majority of the pairs encounter for few number of times, analyzing the encountered pairs without proper grouping could obscure the other significant periodic trends in frequently encountering pairs. To obtain unbiased data, we analyzed the pairs separately according to their encounter rate, more specifically, their daily encounter rate. Result graphs are divided into the two category: rarely encountering pairs and frequently encountering pairs by their daily encounter rate. Let $D_{rate} \left( i,j \right) $ to be a daily encounter rate for a pair $(i,j)$, such that 
\begin{equation}
D_{rate}(x,y)=\frac{\sum_{i=1}^{N}D_{i} (x,y)}{N} 
\end{equation}
This daily encounter rate indicates how many days the pair has encountered over the period of $N$. In the city of Montreal trace, there were no pairs that are $0.1 \leq D_{rate}(x,y)$ due to its scarcely deployed collection devices(APs) in a relatively large city.

\begin{figure*}
\centering
\subfigure[Daily encounter for rarely encountering pairs]{
\includegraphics[width=3in]{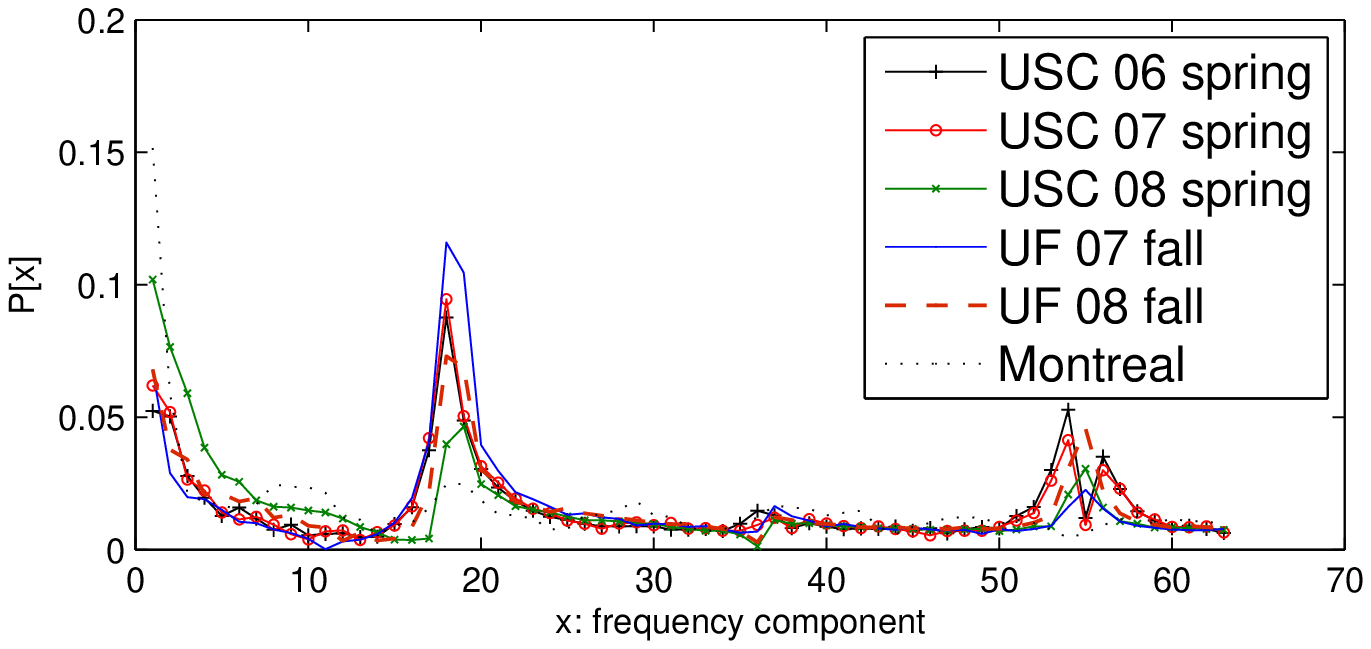} 
\label{daily_rare}
}
\subfigure[Daily encounter for frequently encountering pairs]{
\includegraphics[width=3in]{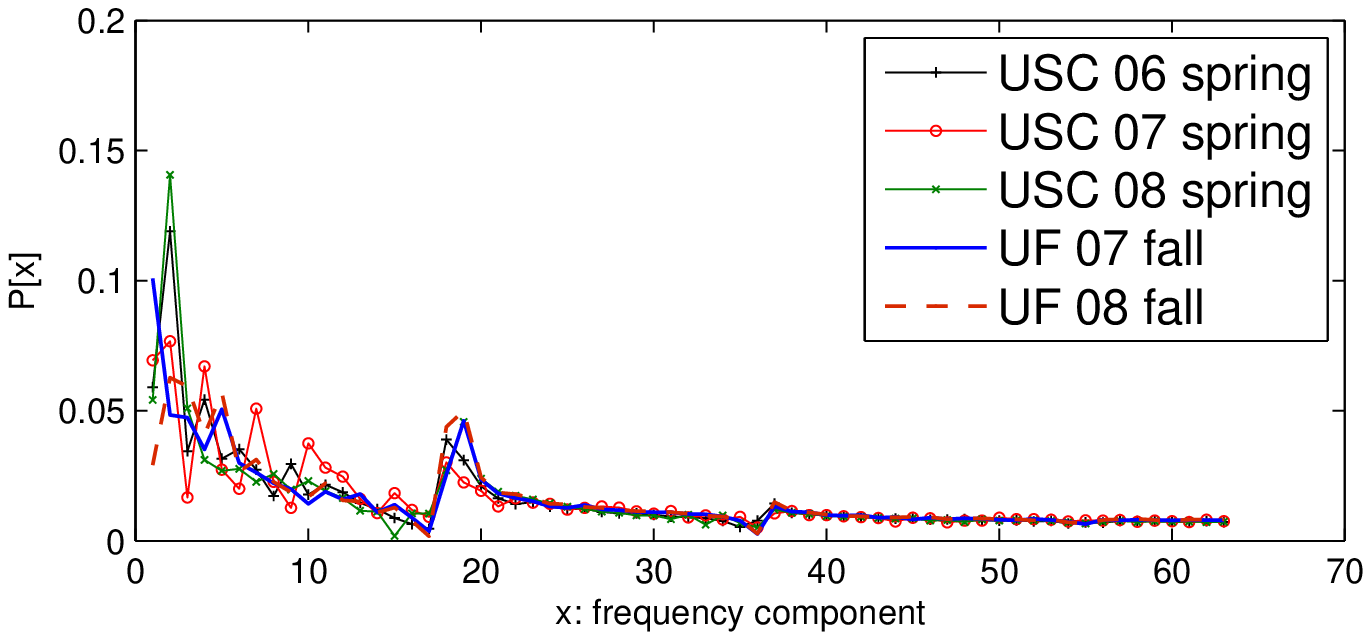} 
\label{daily_frq}
}
\subfigure[Encounter frequency for rarely encountering pairs]{
\includegraphics[width=3in]{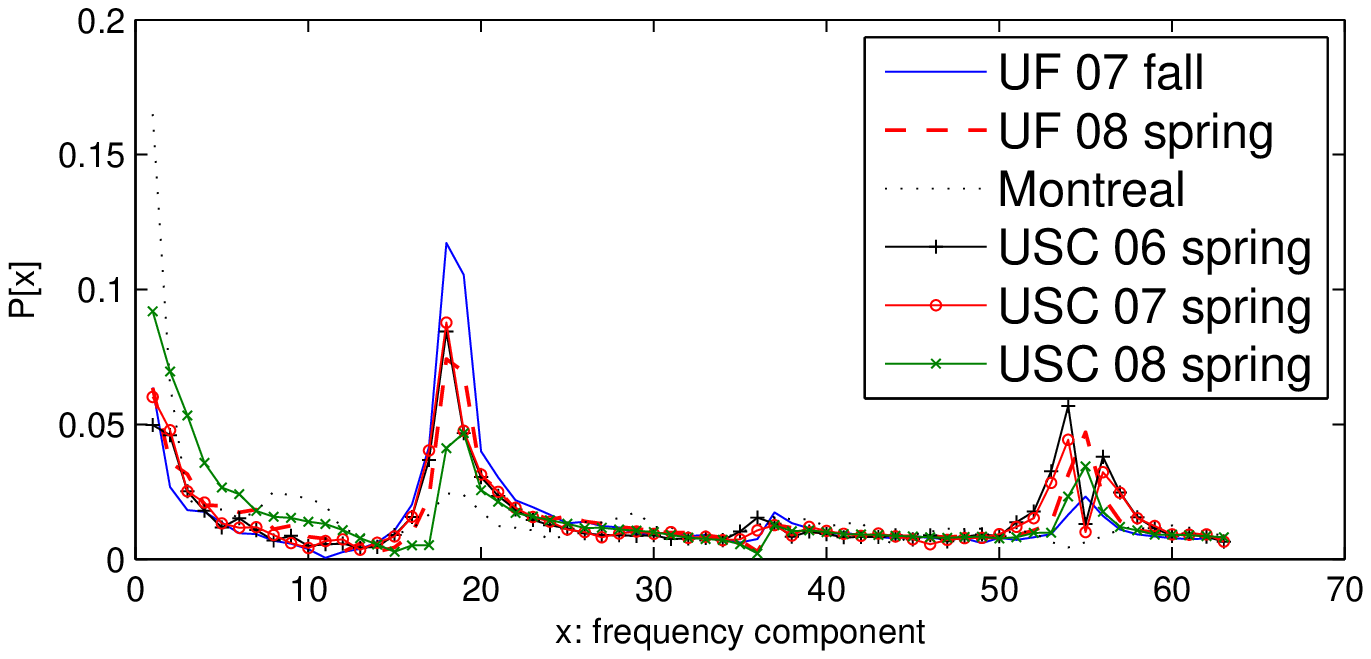} 
\label{frequency_rare}
}
\subfigure[Encounter frequency for frequently encountering pairs]{
\includegraphics[width=3in]{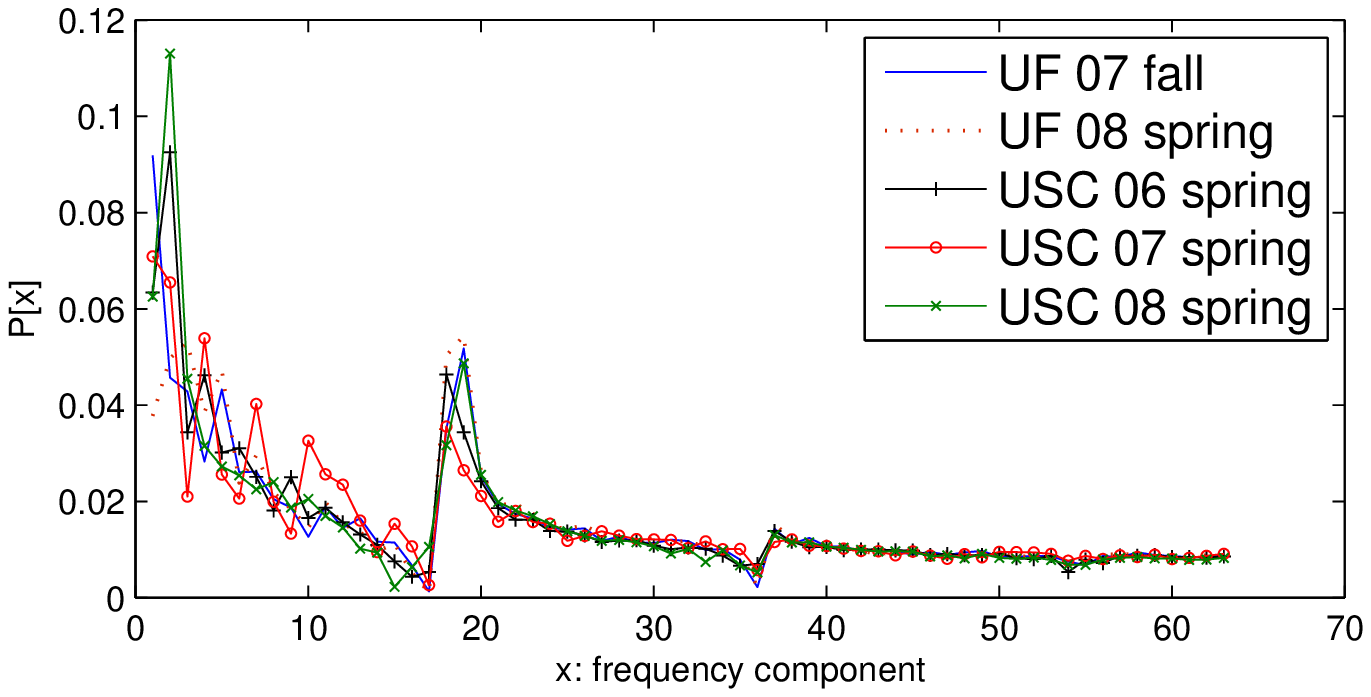} 
\label{frequency_frq}
}
\subfigure[Encounter duration for rarely encountering pairs]{
\includegraphics[width=3in]{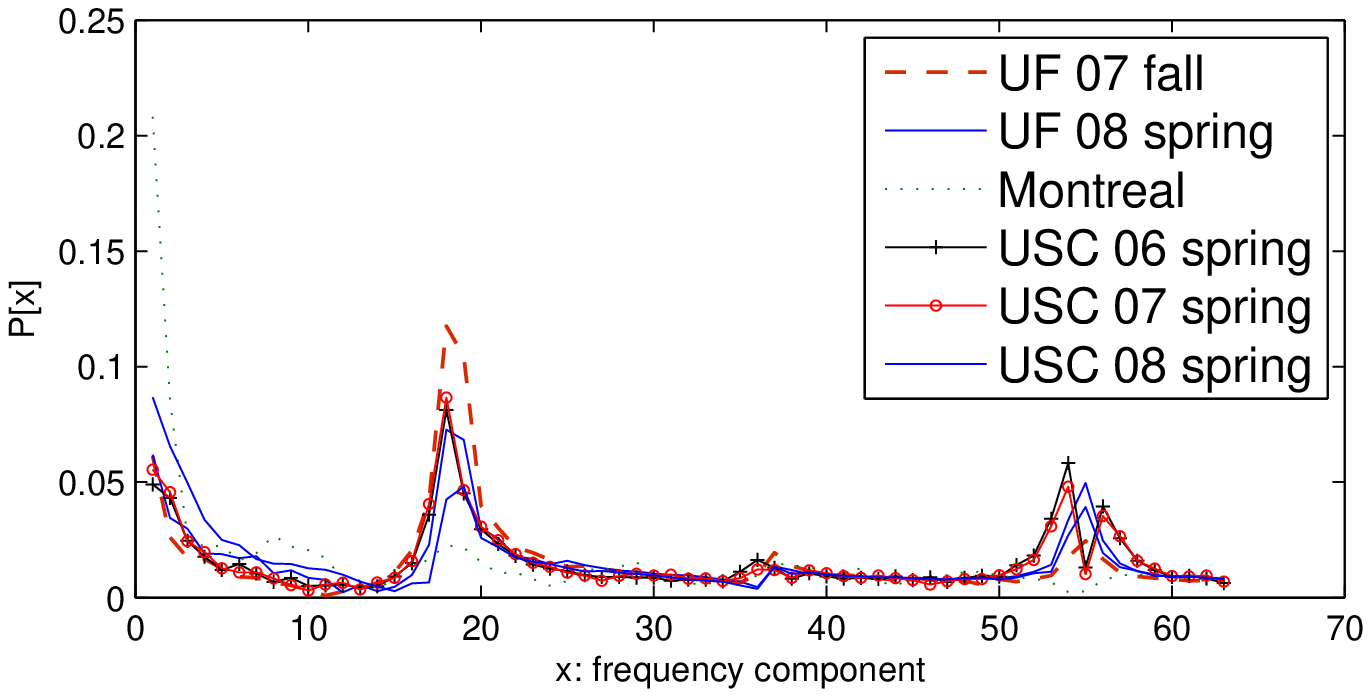} 
\label{duration_rare}
}
\subfigure[Encounter duration for frequently encountering pairs]{
\includegraphics[width=3in]{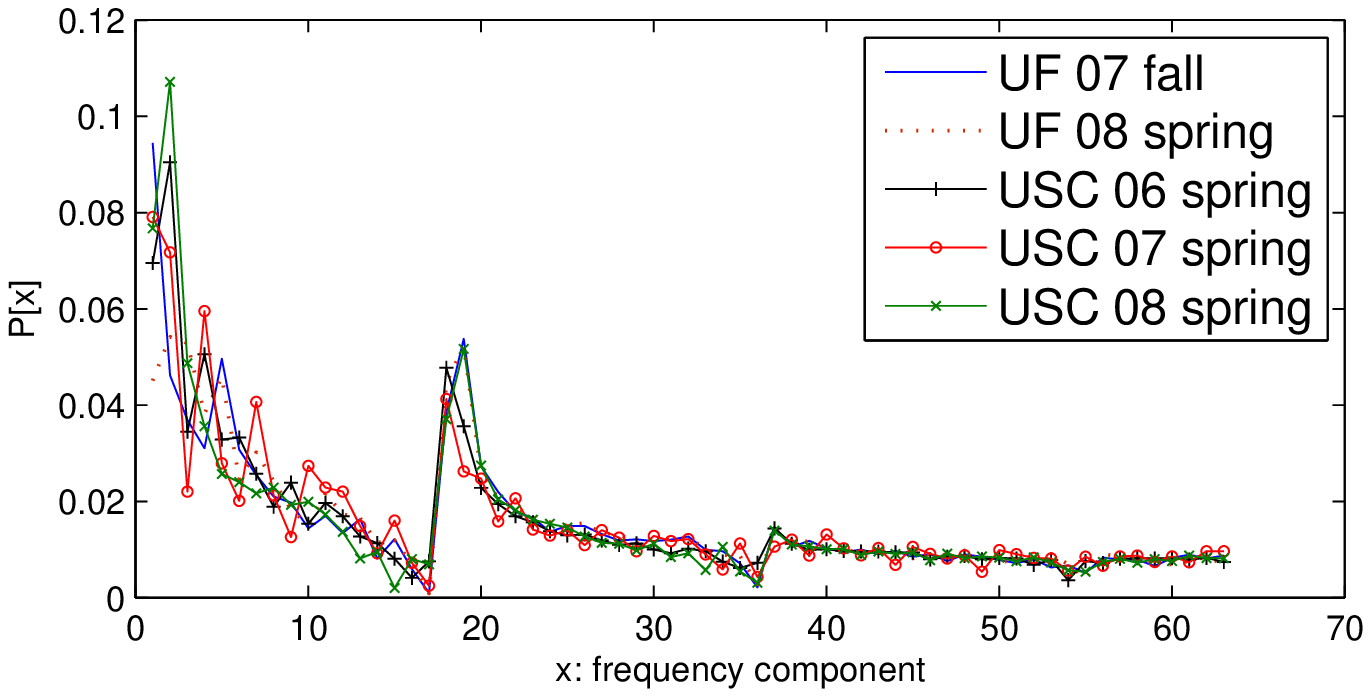} 
\label{duration_frq}
}
\caption{Normalized frequency magnitude of frequency components for encountered pairs $(i,j) (0 \leq i < n, 0 \leq  j < n, i \neq j) $. Rare encouter: $0.1 \leq D_{rate}<0.2$ Frequent encounter: $0.5 \leq D_{rate}<0.6$ (Montreal trace: $0 < D_{rate}<0.1$)}
\label{daily}
\end{figure*}

   From the Fig. \ref{daily}, the large frequency magnitude in axis-Y indicates the strong periodical encounter pattern at the corresponding frequency cycle. It is noticeable that the highest spikes appear at the frequency component of 18, which corresponds to 7 day cycle ($\frac{128}{18} \approx 7.1$) in most of the cases in the Fig. \ref{daily_rare} with an exception of Montreal trace. Fig. \ref{daily_frq} shows that the highest spike appears at the frequency component of 2 for the frequently encountering pairs. This implies that the encounter events may have two big waves but it still shows the presense of 7 day cycle. The presence of strong weekly pattern in encountered pairs is an interesting result as \cite{kim} showed weekly mobility pattern was not among the dominant trends of mobile users' mobility diameter. Consider the logarithmic nature of encounter rate that the large number of pairs encountered less than 20 percent of days over 128 days. Existence of weekly pattern for the pairs with low encounter rate is particularly important in message delivery. Choosing a relay node is a hard decision in a case where majority of nearby nodes encountered infrequently with the delivery target. However, the nodes that show the consistent encounter pattern such as weekly encounter with the target of interest, would likely provide more accurate estimation for delivery probability. With the lower error margin, the source node or intermediate node can further calculate the required number of relay nodes to satisfy the given delivery success rate. Moreover, this implies that the threshold criteria can measured according to the importance of message. Although many other message forwarding schemes can be developed based on the periodic encounter pattern, our focus is on the analysis that can provide more basis for such applications. In Montreal trace, outstanding spikes are hardly shown except in the first frequency, which could suggest the burst encounter pattern but the main reason is the very few number of pairs have encountered repeatedly. Note that there were no pairs, $0.1<D_{rate}$  in Montreal trace. Sparse citywide deployments of the trace collection devices can relate to the low encounter rate among pairs. Besides, low activities and bias in location choices (restaurant, coffee shops) along with spread populations unlike campus environment are thought to be contributing factors. This leads to the question for real world implementation of DTN in large area and it could be an interesting topic to study.

\subsection{Bluetooth trace}

\begin{figure}
\subfigure[Hourly encounter]{
\includegraphics[width=3in]{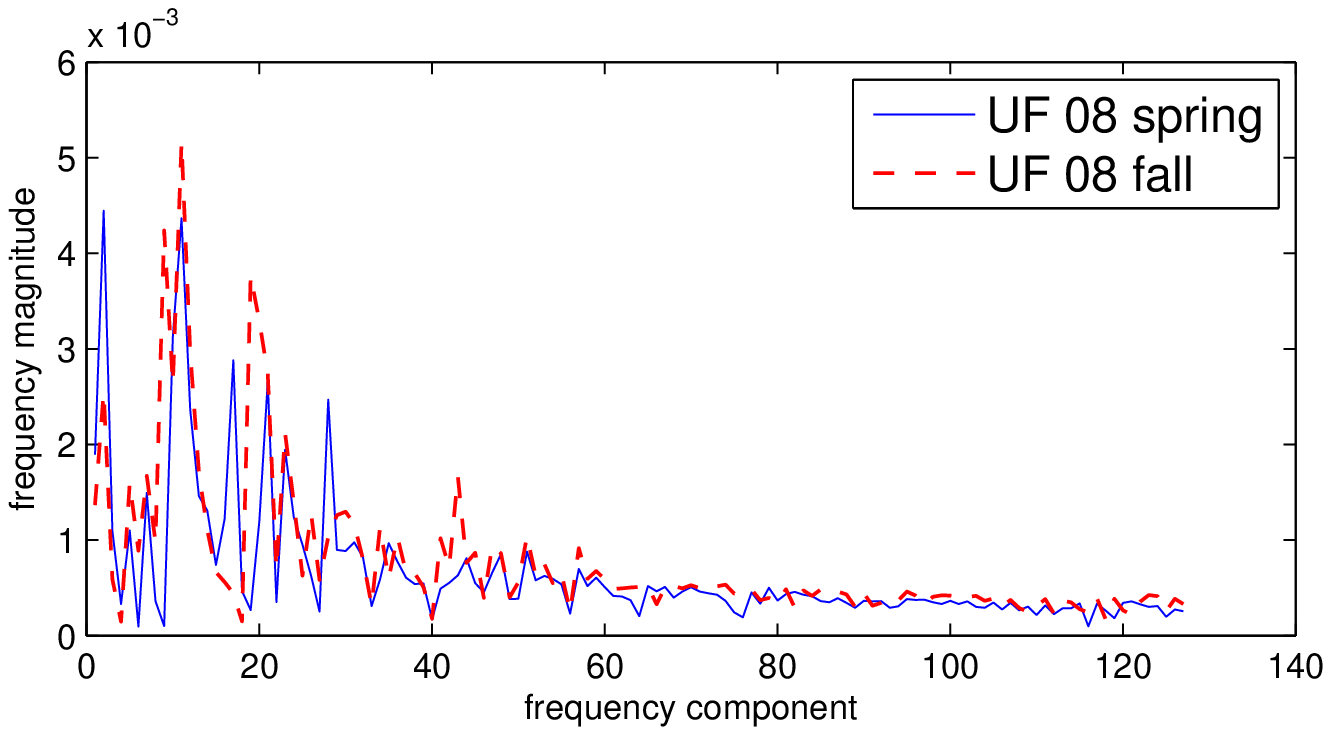} 
\label{bt_uf_hourly}
}
\subfigure[Hourly encounter frequency]{
\includegraphics[width=3in]{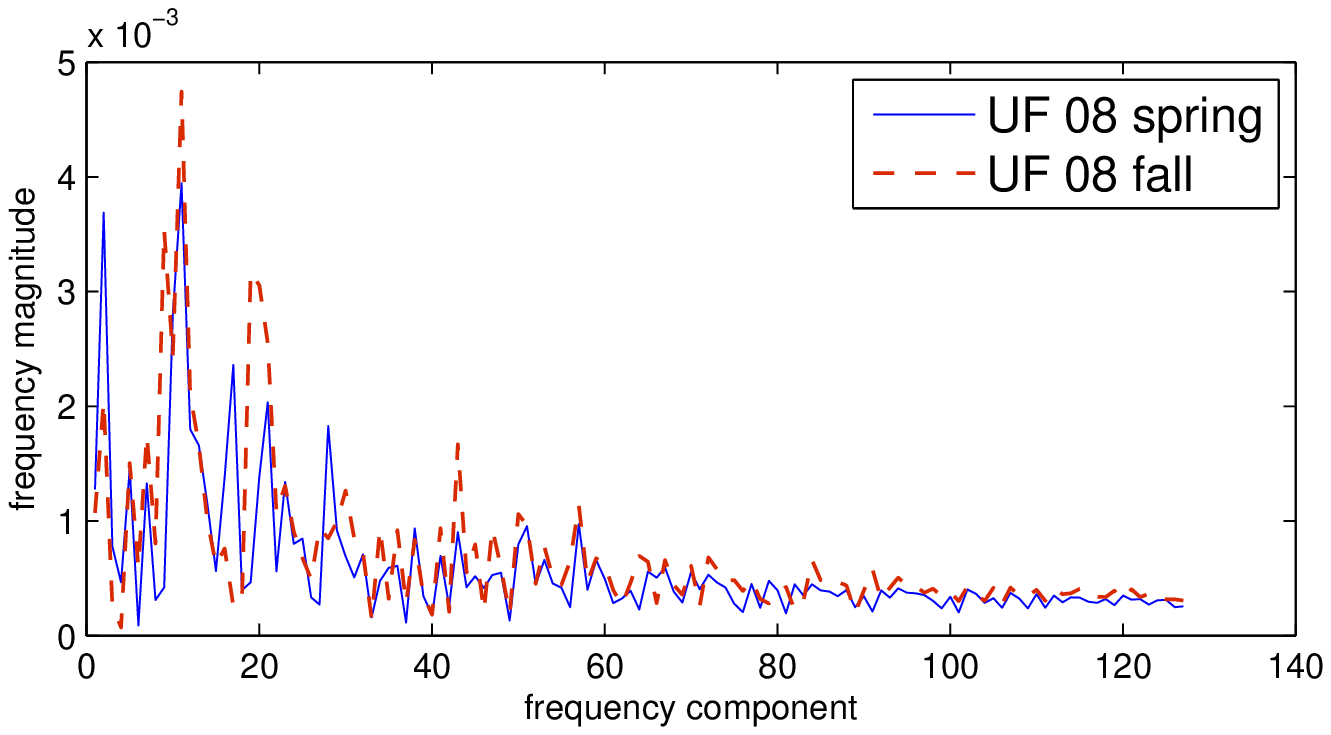} 
\label{bt_uf_freq}
}
\caption{Frequency magnitude of frequency components for hourly encounter at UF 08 spring/fall Bluetooth trace}
\label{bt_uf}
\end{figure}

   We study the hourly encounter for bluetooth experiment due to the short length of the bluetooth experiment. The difference from observing daily encounter is granularity of observation is finer. In this experiement, we look at the 256 hours, which is approximately 10 days. Fig. \ref{bt_uf_hourly} shows the hourly encounter patterns of encountered pairs with $ 0.2 \leq D_{rate} < 0.3 $. In the Figure, axis-X indicates the frequency of cycles for the experiment period in hours. Given $D_{rate}$ was selected because $0.1 \leq D_{rate} < 0.2$ was unavailable due to experiment length limitation. Accoring to the figure, 24-hour periodicity is strongest in both of the Bluetooth encounter traces. Hourly encounter frequency in Fig \ref{bt_uf_freq} displays more periodicitic pattern but it is still similar to hourly encounter. The graphs indicate periodic encounter occurs around every 24 hour in average; thus, suggest that most of the encounter events may occur during the similar time span of the day. This 24 hour periodicity in encounter pattern corresponds to the result in mobility diameter study in \cite{kim}.

\subsection{Periodicity of individual encounter pattern}

\begin{figure*}
\centering
\subfigure[Daily encounter for rarely encountering pairs]{
\includegraphics[width=3in]{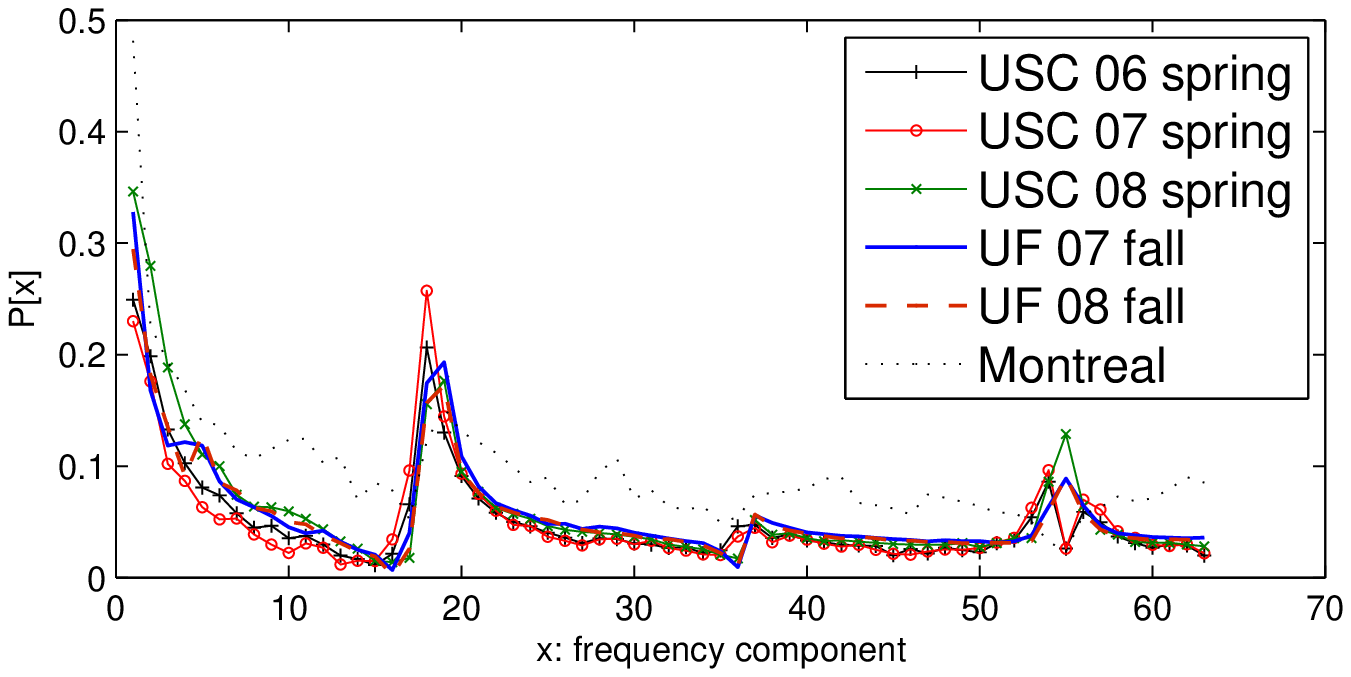} 
\label{ind_rare_all}
}
\subfigure[Daily encounter for frequently encountering pairs]{
\includegraphics[width=3in]{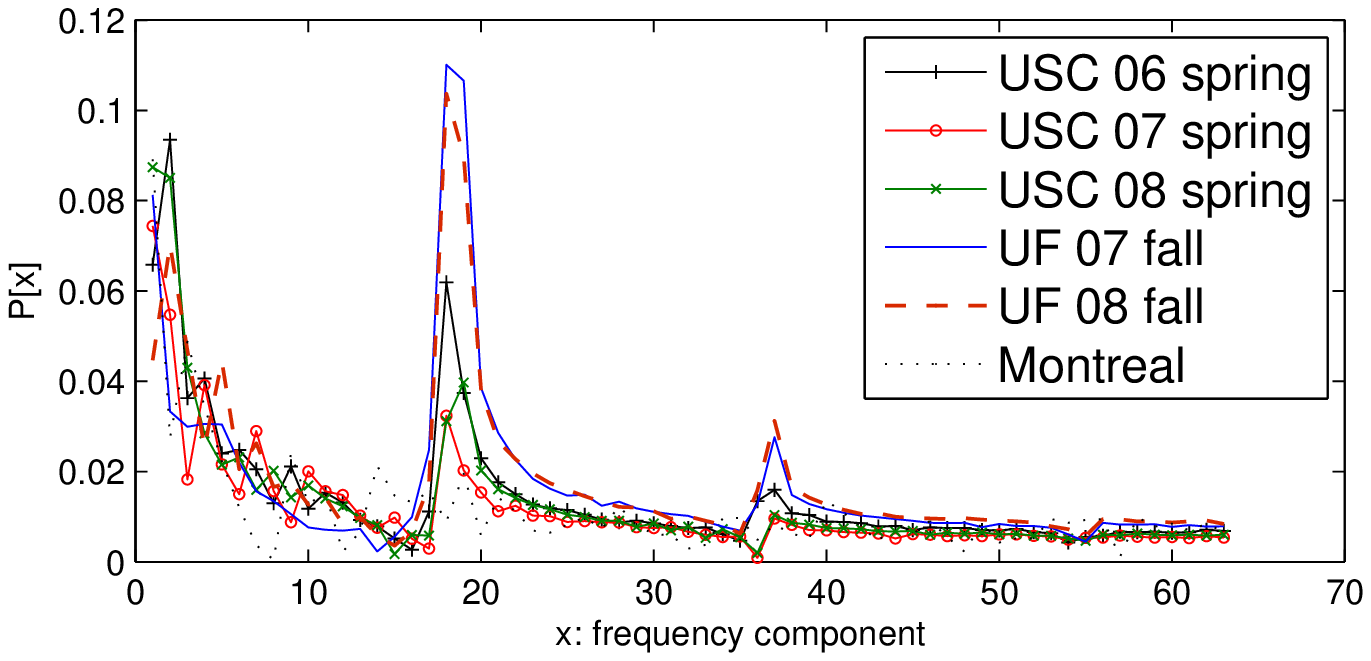} 
\label{ind_frq_all}
}
\subfigure[Encounter frequency for rarely encountering pairs]{
\includegraphics[width=3in]{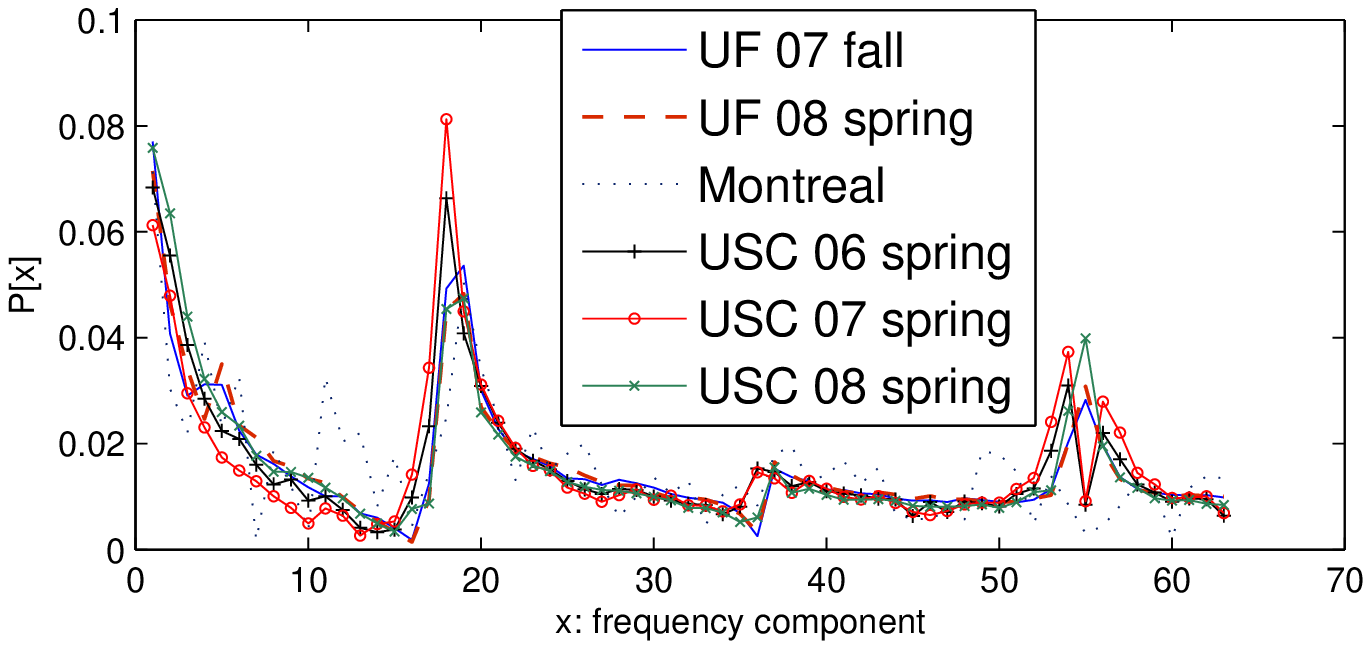} 
\label{ind_rare_all}
}
\subfigure[Encounter frequency for frequently encountering pairs]{
\includegraphics[width=3in]{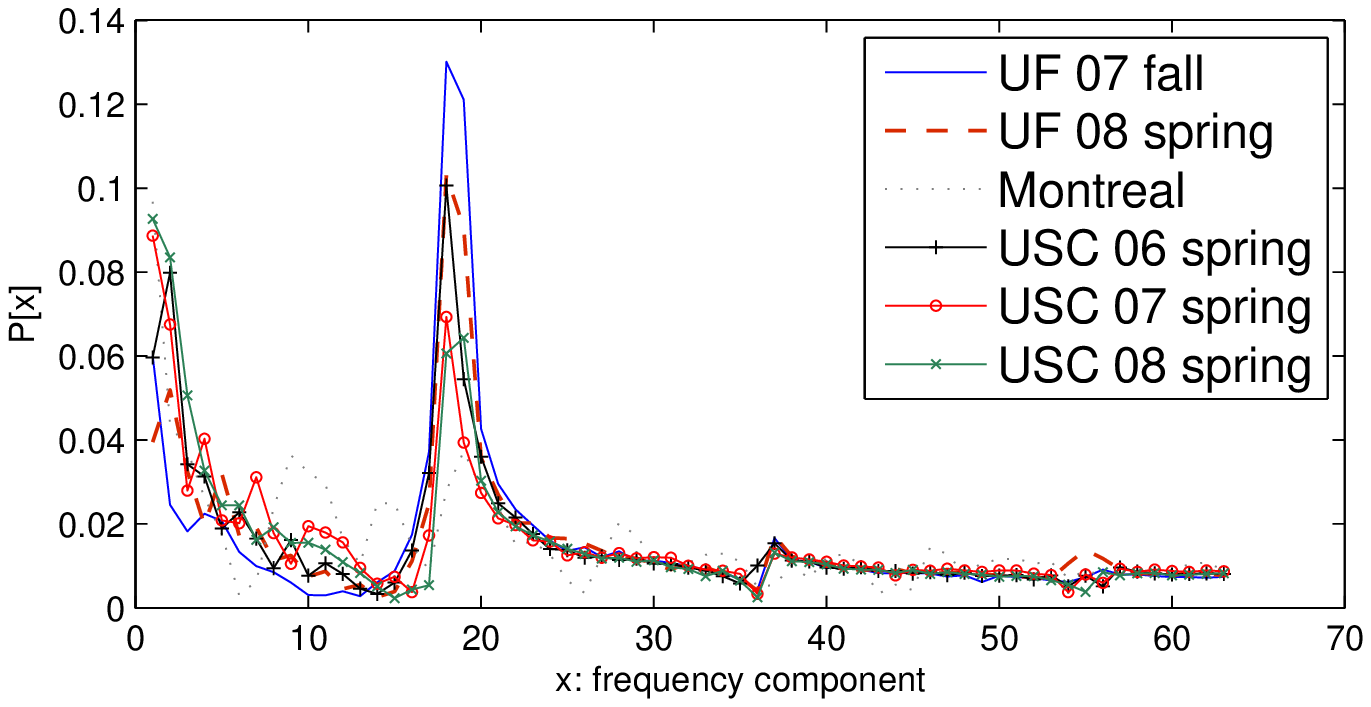} 
\label{ind_rare_all}
}
\subfigure[Encounter duration for rarely encountering pairs]{
\includegraphics[width=3in]{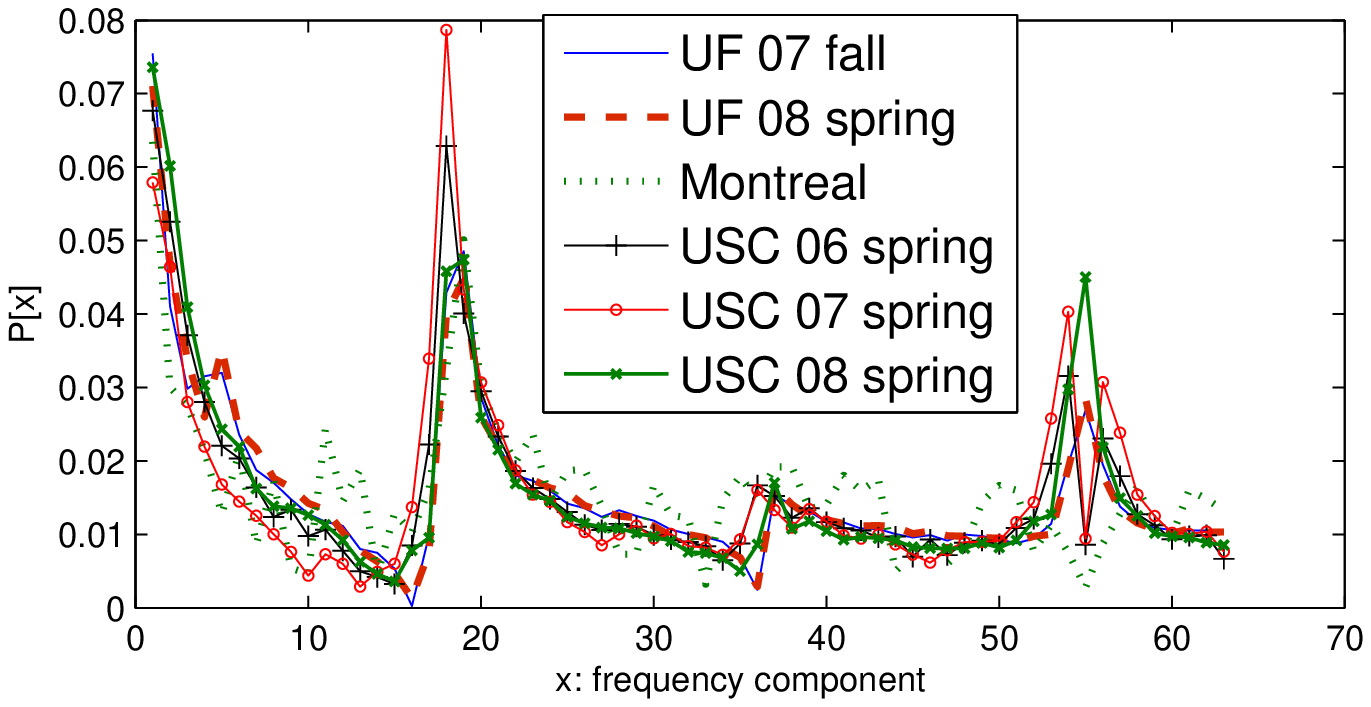} 
\label{ind_rare_all}
}
\subfigure[Encounter duration for frequently encountering pairs]{
\includegraphics[width=3in]{newfigures/fft_eachnode_duration_rare_all.eps} 
\label{ind_rare_all}
}
\caption{Normalized frequency magnitude of frequency components for individual nodes' encounter pattern. Rare encouter: $0.1 \leq D_{rate}<0.2$ Frequent encounter: $0.5 \leq D_{rate}<0.6$ (Frequent encounter of Montreal trace: $0.2 leq D_{rate}$)}
\label{ind_all}
\end{figure*}

   The periodicity in the encounter pattern of individual node is even stronger than in the encounter pattern of pairs. Let $D_{rate}(i)$ be a daily encounter rate for a node $i$, such that $D_{rate}(i)=\frac{\sum_{d=1}^{N}E_{d}(i)}{N}$, where $E_{d}(i)$ is $1$ if at least one encounter event occurred on the day i and $0$ otherwise. Note that peak for weekly encounter in encounter pattern of each node in Fig. \ref{ind_rare_all} is more distinct than the peak in each encounter pair in Fig. \ref{ind_frq_all}. This implies that aggegate encounter behavior of each node is more periodical than the encounter behavior of pairs; thus, consistently more predictable. However, for the purpose of message delivery, understanding the encounter behavior of pairs is more useful than studying the encounter behavior of individual node. This is because the source node will make a decision of selecting a relay node based on the information about encounter behavior of the candidate node to the destination node rather than its overall encounter behavior with all nodes. In the figure, periodic encounter pattern is most prevalent and consistent in the sequency of daily encounter, encounter frequency and encounter duration at both encounter patterns in the pairs and individual nodes. We also observed periodic encounter occurrence at every 24 hour from hourly encounter pattern in both of the WLAN and bluetooth traces for individual encounter. Naturally, we can infer that the nodes are penchant to encounter in similar hours of the day each day. Note that in the figure, as the shape of the concave is wider, periodic nature is less accurate or has wider margin of error. The narrower and higher the bell shape of power is, the stronger the periodic property is.

\section{Regular encounter}
   To utilize the periodic properties of encounter pairs, it is essential to develop a scheme to discover such pairs in success of forming a network among those nodes. With the transformed data in frequency domain, it is simple to extract the pairs that encounter consistently in a periodic fashion, which we define as regularly encountering pairs. The challenge in regularity is that it can have multiple variables in it. Several trends can be hidden and noise may interfere from observing some of the periodicities. We discuss the several approaches that extract regularity from the traces and present the result for USC'06 trace.

\subsection{Top frequency}

\begin{figure}
\includegraphics[width=3in]{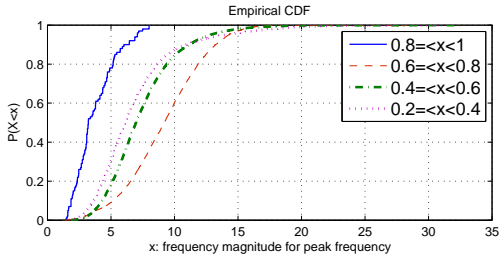} 
\caption{Empirical CDF of highest frequency for the encountered pairs at USC 06 spring trace according to daily encounter rate (x = $ D_{rate} $ )}
\label{top_cdf_usc06}
\end{figure}

\begin{figure}
\includegraphics[width=3in]{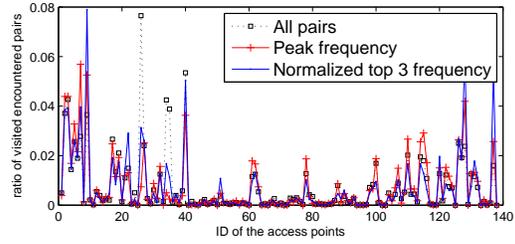} 
\caption{APs accessing preference at USC 06 spring trace for $ 40 \leq D_{rate} < 60 $}
\label{location_4060}
\end{figure}

   The large frequency magnitude at the left most frequency in axis-X suggests that the encounter events are likely occurring at a concentrated time, forming one big wave. Conversely, the large frequency magnitudes at high frequencies in axis-X indicate the more waves existed in the time domain representation, closer to uniform distribution. Therefore, the pairs with strong trend in high frequencies are search of interests - regularly encountering pairs. Although the notable weekly pattern was showing in average, periodicity appears differently by each pair. Fig. \ref{top_cdf_usc06} shows cdf of the highest frequency of the encountered pairs in order of frequency magnitude. Right knee in the upper right side of the graph indicates that periodic behavior is stronger for some of the pairs, suggesting regular encounter activity. Based on this observation, we grouped the pairs that encounter regularly by taking the pairs whose top frequencies are over the knee point(top 20 percent). We plot the APs where the regularly encountering pairs have accessed and overall pairs have accessed in Fig. \ref{location_4060}. It is clear that the location visiting patterns at the time of encounters are notably different in many of the locations. Note that the lowest frequency was not considered in grouping the regularly encountering nodes for the reason that it does not indicate the regular encounter, rather burst encounter. 

\subsection{Normalized regular encounter}

\begin{figure}
\centering
\includegraphics[width=3in]{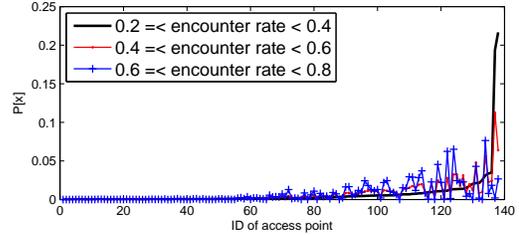} 
\caption{Ordered location visiting preference by encountered pairs according to daily encounter rate}
\label{ordered_loc_overall_pairs}
\end{figure}

   Taking only the top frequency magnitude may not capture the regularly encountering pairs accurately, because the number of spikes in the frequency graph can be of two or more. This is normal for regular pattern as some of the trend can consist of several minors including artifacts of frequency analysis (e.g. cycles at every 8 time unit can cause to have cycles at every 16 time unit). To better capture the regularly encountering pattern while considering the possible noise factors, we observed normalized top frequencies for each pair. To achieve the normalization, the summation of several top frequency magnitudes was compared to the summation of all the frequency magnitudes in each pair. We observed that the top 3 frequencies were taking up more than one third of all for the most of the pairs consistently across the different daily encounter rate. Hence, we use the top three frequencies as a criteria to extract the regularly encountering pairs. After applying this rule, it appears that the ratio of regularly encountering pairs to the overall pairs differ by the daily encounter rate with the range being from 0.15 to 0.4. Rare encounter can hardly yield regularity due to its small number of samples. For the opposite reason, frequent encounter often leads to the uniform distribution of encounter, thus, any patterns are hardly noticable especially in daily encounter.

   We first discuss the general trend of location visiting pattern at the time of encounter events. Then, we compare the result to the regularly encountering pairs. In Fig. \ref{ordered_loc_overall_pairs}, the data is sorted according to the number of encounter events in the location. It is clear from the figure that encounter events are skewed to a few locations, thus, the graph showing exponential curve toward highly visited locations. Another observation is that frequently encountering pairs are exhibiting different location visiting patterns from less frequent pairs at the time of encounter events. This implies that the locations with rare encounter events are disadvantaged in message delivery. Now we turn our attention to the trend for the regularly encountering pairs. In the Fig. \ref{location_4060}, the regularly encountering pairs are showing different visiting preference. This different location visiting pattern supports the strong value of using regular encounter pattern. Delivery attempt to the nodes that mainly appear in the location of scarce encounter events may suffer from finding the forwarding nodes that have history of encounter event. Therefore, overload on the network can be expected due to too many number of message copies. In such a situation, 1) using the regularly encountering nodes if the encounter rate is similar, would make more accurate estimation for the number of relay nodes to reach the delivery probability goal and/or 2) the source node has a more chance to discover the nodes that regularly encounter with the target node in some locations. More applications can be developed that use such characteristics; thus, our analysis opens up for more applications and is one of our contributions. 

\section{Related work}
   Many of studies on DTN/opportunistic/intermittent connectivity routing were devoted in using the social aspects of networks, such as community, mobility and encounter. Gonzalez et al. \cite{mobilitynature} has shown that individual human tends to follow simple reproducible patterns based on the cell phone user traces. Hsu et al. proposed time-variant community model \cite{timevariant} that reflects the periodic encounters. Social relationship between mobile nodes were discussed for DTN routing in \cite{dalysocial} \cite{bubblerap}. Miklas et al. \cite{miklassocial} divided human encounters to friends and strangers according to the length of encounter. These works study human encounter pattern; however, we are the first to analyze the periodicity of human encounter extensively by spectral analysis.  Prophet \cite{prophet} is one of the first routing algorithms using encounter history in DTN \cite{dtn}. It uses encounter frequency to determine a relay node. The chosen node will forward the message bundle to the encountered node and delegate the responsibility of delivery to the node if the node has higher encounter frequency to the destination node. However, it is unclear how to determine the probability of encounter using frequency. To use encounter frequency for probability, total number of possible encounters should be known, which otherwise could be infinite. Further, it is possible that frequent encounters in short time can mislead prediction of future encounter. Timely-count probability \cite{interval} is an idea that regards encounters belonging to the same interval as one encounter; thus, it provides standard procedure to calculate the encounter probability. In our work, we used it for an idea of daily encounter whose interval is a day. 
   Our periodicity and regularity analysis can be the basis to improve the performance of protocols in DTN. Profile-cast \cite{profilecast} is a forwarding protocol to the group of nodes, sharing the same interest. Both profile-cast and prophet protocols can enhance their stability of delivery by incorporating periodic properties of encounter for consistent predictability. Studies for predictability of human mobility and encounter \cite{songlimit} \cite{huipredict} can also be extended by considering different periodic properties as well as worm propagation pattern via human encounter \cite{sapon} \cite{wangvirus}. 
   \cite{kim} studied the periodic properties of WLAN users¡¯ association with access points. They measured the diameter of visited APs for highly mobile users whose maximum diameter within an hour is 100 meters or more from the Dartmouth campus WLAN data. The result showed strong presence of periodicity of diameter, particularly 24 hours for the selected 360 users. This is the closest work to our frequency analysis in that it uses DFT on the time series data to analyze the periodicity of user mobility. \cite{attack} analyzed the network traffic and revealed the presence of combination of periodicity in the case of denial of service attacks on the internet. They applied power spectral analysis that applies ACF to the time series data before transforming to frequency domain. This removes sync terms that might appear for a finite set, therefore, we adapted a similar approach rather than applying DFT to the data sets directly.

\section{Conclusions}
   We analzed the periodicity in encountered pairs and individual nodes under various conditions. We categorized them according to their daily encounter rate and showed the periodicity with the following metrics: daily and hourly encounter, encounter frequency and encounter duration. For the majority of the encountered pairs, a weekly encounter pattern was prevalent, which was not previously observed in mobility diameter pattern. We also observed that periodicity appeared stronger for the rarely encountering pairs than the frequently encountering pairs. In case of rare encounter events, the regular encounter pattern is particularly useful as it can provide the estimation for the number of required relay nodes to satisfy the given delivery probability. We also proposed viable approaches to discover the regularly encountering pairs. Our analysis shows the utility of spectral analysis for characterizing encounter regularity, which is vital for future mobile networks. Additionally, we showed that regularly encountering pairs may have different location visiting patterns, which further reinforces the importance of regular encounter pattern. We analyzed the real-world encounter data sets (Bluetooth encounter) and the WLAN traces with adequate assumption for large-scale encounter data, where periodicity in various setup was commonly observed at both types of traces. To sum up, our analysis shows the utility of spectral analysis for characterizing encounter regularity, which is vital for the study of future mobile networks. Our work is unique in that we 1) investigate the periodicity and regularity of nodal encounter pattern by using power spectral analysis 2) propose new approaches to discover the regular pattern and analyze the regularly encountering pairs 3) analyze a rich set of real world data for long periods of time, including up to 50,000 users per semester period over three years and Bluetooth traces in two semesters.

\bibliographystyle{IEEEtran}

\bibliography{globecom_sungwook}

\end{document}